\documentclass[pra,nobalancelastpage,twocolumn,nofootinbib,superscriptaddress,longbibliography]{revtex4-1} 
\pdfoutput=1

\usepackage{color,amsthm,amsmath,amsxtra,amsfonts,dsfont,graphicx,bm}
\usepackage[section]{placeins}

\newcommand\nn{\mathbf{n}}
\newcommand\ii{\mathbf{i}}
\newcommand\jj{\mathbf{j}}

\newcommand\kk{\mathbf{k}}

\newcommand\rr{\mathbf{r}}

\newcommand\as{\mathcal{A}}
\newcommand\bs{\mathcal{B}}

\newcommand{\bra}[1]{\left\langle #1\right|}
\newcommand{\ket}[1]{\left| #1\right\rangle}

\newcommand{\braket}[2]{\langle #1|#2\rangle}
\newcommand{\bkev}[3]{\langle #1|#2| #3 \rangle}
\newcommand{\pa}[1]{\left( #1\right)}
\newcommand{\ha}{H}

\newcommand{\abs}[1]{\left| #1\right|}

\newcommand{\fl}[1]{\left\lfloor #1\right\rfloor}
\newcommand{\cl}[1]{\left\lceil #1\right\rceil}
\newcommand{\mean}[1]{\left \langle #1 \right \rangle}

\usepackage{accents}

\usepackage{hyperref}
\usepackage[margin={0pt,0pt}, font=small,labelfont=bf,justification=centerlast]{caption}

\definecolor{mygrey}{gray}{0.35}
\definecolor{myblue}{rgb}{0.2,0.2,0.8}
\definecolor{myzard}{cmyk}{0,0,0.05,0}
\definecolor{mywhite}{rgb}{1,1,1}
\definecolor{mywhite}{rgb}{1,1,1}
\definecolor{myred}{rgb}{1,0.,0.3}

\newcommand{\norm}[1]{\left\lVert#1\right\rVert}
\newcommand{\co}[1]{\left[ #1\right]}
\newcommand{\MPQ}{Max-Planck-Institut f{\"u}r Quantenoptik, Hans-Kopfermann-Stra{\ss}e\ 1, D-85748 Garching, Germany}

\hypersetup{
  pdfsubject={},
  pdfcreator={pdflatex},
  colorlinks,
  linkcolor=blue,
  citecolor=blue,
  urlcolor=blue
}

\usepackage{appendix}
\usepackage{booktabs,array,multirow}

\usepackage{subfig}
\usepackage{ bbold }

\newcommand{\be}{\begin{equation}}
\newcommand\bb[1]{\mathbf{#1}}
\newcommand{\ee}{\end{equation}}
\newcommand{\bea}{\begin{eqnarray}}
\newcommand{\eea}{\end{eqnarray}}

\begin{document}

\title{Analog quantum chemistry simulation}
\date{\today }

\begin{abstract}
Computing the electronic structure of molecules with high precision is a central challenge in the field of quantum chemistry. Despite the enormous success of approximate methods, tackling this problem exactly with conventional computers is still a formidable task. This has triggered several theoretical~\cite{aspuru05a,Wecker2014a} and experimental~\cite{OMalley2016,Kandala2017,Lanyon2010} efforts to use quantum computers to solve chemistry problems, with first proof-of-principle realizations done in a digital manner. An appealing alternative to the digital approach is analog quantum simulation, which does not require a scalable quantum computer, and has already been successfully applied in condensed matter physics problems~\cite{choi16a,trotzky12a,Greiner2002}. However, all available or planned setups cannot be used in quantum chemistry problems, since it is not known how to engineer the required Coulomb interactions with them. Here, we present a new approach to the simulation of quantum chemistry problems in an analog way. Our method relies on the careful combination of two technologies: ultra-cold atoms in optical lattices and cavity QED. In the proposed simulator, fermionic atoms hopping in an optical potential play the role of electrons, additional optical potentials provide the nuclear attraction, and a single spin excitation over a Mott insulator mediates the electronic Coulomb repulsion with the help of a cavity mode. We also provide the operational conditions of the simulator and benchmark it with a simple molecule. Our work opens up the possibility of efficiently computing electronic structures of molecules with analog quantum simulation. 
\end{abstract}

\author{Javier Arg\"uello-Luengo}
\affiliation{\MPQ}
\affiliation{Institut de Ciències Fotòniques (ICFO), The Barcelona Institute of Science and Technology, 08860 Castelldefels, Spain}

\author{Alejandro Gonz\'alez-Tudela}
\email{alejandro.gonzalez-tudela@mpq.mpg.de}
\affiliation{\MPQ}
\affiliation{Instituto de F\'isica Fundamental IFF-CSIC, Calle Serrano 113b, Madrid 28006, Spain}

\author{Tao Shi}
\affiliation{\MPQ}
\affiliation{CAS Key Laboratory of Theoretical Physics, Institute of Theoretical Physics, Chinese Academy of Sciences, P.O. Box 2735, Beijing 100190, China}

\author{Peter Zoller}
\affiliation{Center for Quantum Physics, University of Innsbruck, A-6020 Innsbruck, Austria}
\affiliation{\MPQ}

\author{J. Ignacio Cirac}
\email{ignacio.cirac@mpq.mpg.de}
\affiliation{\MPQ}
\affiliation{Munich Center for Quantum Science and Technology (MCQST), Munich, Germany}
\maketitle


\begin{figure*}[!htb]
\centering
\includegraphics[width=0.8\linewidth]{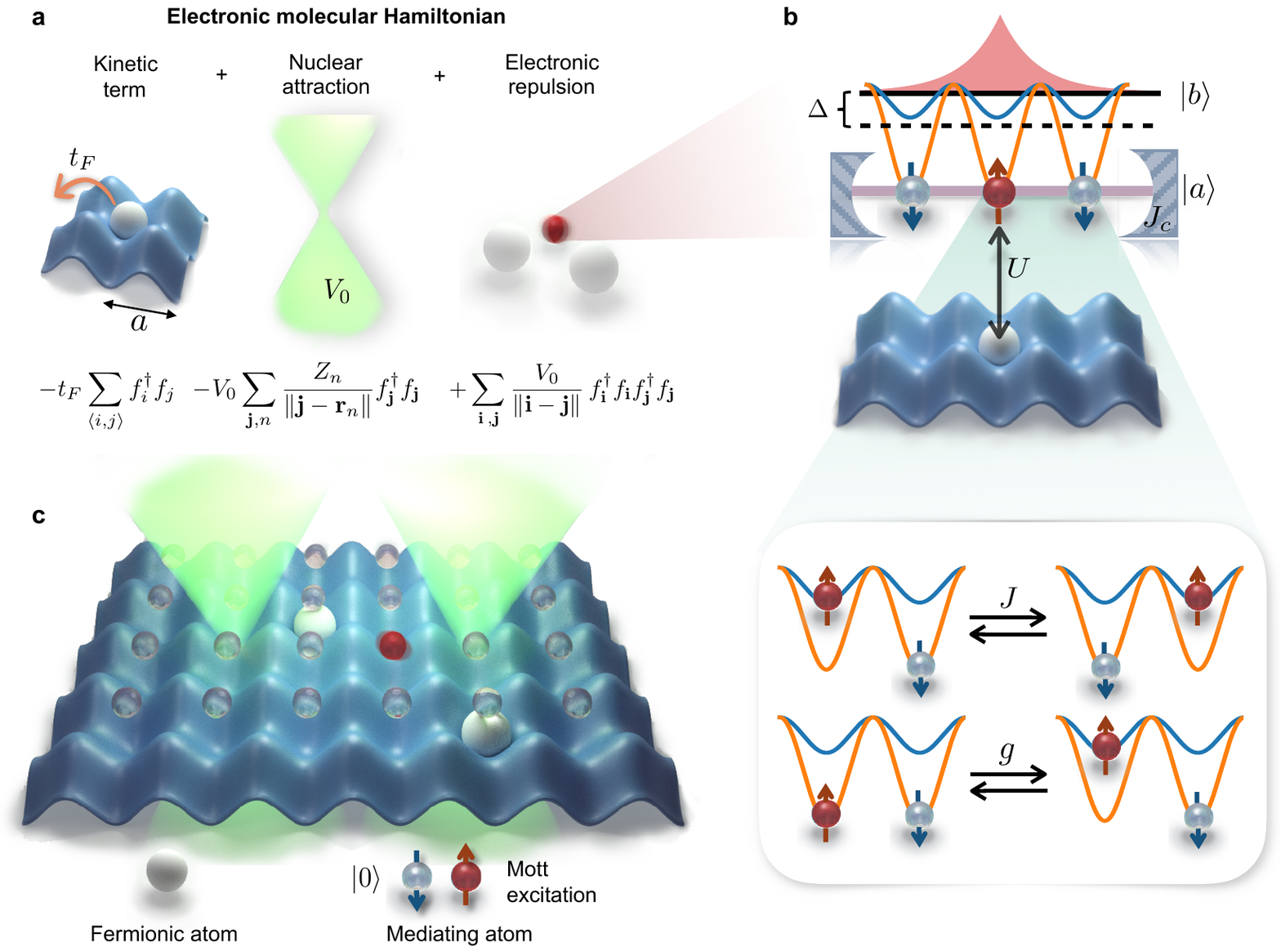}
\captionsetup{labelformat=empty}
\caption*{\textbf{Fig. 1 $|$ Schematic representation of the analog simulator.} \textbf{a,} Fermionic atoms, playing the role of electrons, are trapped in a periodic 3D cubic potential. Their hopping simulates the kinetic energy term of the electrons, and they are subject to additional optical potentials which emulate the nuclear interaction.  \textbf{b,} Coulomb repulsion among the fermions is mediated by a spin excitation of a Mott insulator with three internal levels. Excitations in level $|b\rangle$ are allowed to propagate through spin exchange interactions with rate $J$ (upper in the box). Excitations in level $|a\rangle$ experience a strong repulsive interaction with the fermions and interacts with a cavity mode. These two levels are coupled either through a microwave or a two-photon Raman transition (lower in the box).  \textbf{c,} Illustration of the complete simulator for the $H_2$ molecule. While a bidimensional lattice is pictured, the experimental proposal presented here refers to a three-dimensional optical lattice.}
\end{figure*}

Quantum computers are expected to have a significant impact in several areas of science, as they will be able to tackle problems which are intractable with classical devices. Particularly relevant are quantum many-body problems involving several systems interacting with each other according to the rules of Quantum Physics \cite{Feynman1982}. Among the most timely and significant ones are quantum chemistry problems, which generally imply obtaining the ground state energy of many electrons interacting with both the nuclei and among themselves through Coulomb interactions. Current approaches to the simulation of chemistry problems with quantum computers follow the digital approach~\cite{Cao2019,McArdle2020,Kassal2011,Reiher2017}, in which one breaks the complete Hamiltonian in gates that are applied in a time dependent manner.

An alternative way to address quantum many-body problems is analog quantum simulation~\cite{cirac12a}. The idea is to use a well-controlled quantum system (the simulator) and engineer its interactions according to the Hamiltonian under investigation. This approach has already addressed questions that the most advanced classical computers cannot compute~\cite{choi16a,trotzky12a,Greiner2002}. The key feature is that their interactions are either local or short-range, being ideally suited for the existing simulators. On the contrary, analog simulation of quantum chemistry requires engineering long-range (Coulomb) interactions between fermionic particles, and no system has been identified so far fulfilling such requirement. This is why current efforts concentrate in digital simulation.

In this work we show how to build an analog simulator for quantum chemistry problems by bridging two paradigmatic systems, namely, ultra-cold atoms in optical lattices~\cite{bloch08a,Murch2008,Krinner2018} and cavity QED~\cite{ritsch13a,Schreppler2014,Brennecke2008,Brennecke2007,Domokos2002}. Fermionic atoms trapped in a periodic 3D optical potential play the role of electrons and are subject to additional optical potentials emulating their interaction with the nuclei. The key ingredient of the scheme is to trap another atomic species in the Mott insulator regime, with several internal states such that its spin excitations mediate effective forces between the simulated electrons. We show that even though the interaction is local, one can induce Coulomb-like forces among the fermionic atoms in a scalable manner. While the setup is discrete and finite, we show that precise results can be obtained for simple molecules with moderate 
lattice sizes. Apart from the standard advantages of analog simulation over quantum computing regarding the required control~\cite{cirac12a}, the present scheme does not rely on a judicious choice of molecular orbitals~\cite{szabo2012modern}, but directly operates in real space improving convergence to the exact result as the system size increases.

One of the main goals of Quantum Chemistry is to obtain the low energy behavior of $N_e$ electrons and several nuclei when the positions, $\rr_n$, of the nuclei are fixed. Using a cubic discretization in real space of $N\times N\times N$ sites, the electronic Hamiltonian to solve contains three terms, $H_\mathrm{qc}=H_\mathrm{kin}+H_\mathrm{nuc}+H_{e-e}$, (using $\hbar=1$, and dropping the spin index) 
\begin{align}
\label{eq:proyHam}
& H_\mathrm{kin} = -t_F\,\sum_{\langle \ii,\,\jj \rangle}f^\dagger_\ii f_ \jj\,,\\ & H_\mathrm{nuc}= -\sum_{n,\,\jj} Z_n V(\norm{\jj-\rr_n})\,f^\dagger_\jj f_ \jj \\  & H_{e-e}= \sum_{\ii,\,\jj} V(\norm{\ii-\jj})\,f^\dagger_\ii f_ \ii f^\dagger_\jj f_\jj\,,
\end{align}
where $f_\ii$ are annihilation operators of electrons at site $\ii$ fulfilling $\{f_\ii,f^\dagger_\jj\}=\delta_{\ii,\jj}$, and $\langle \ii,\,\jj \rangle$ denote nearest neighbour sites. $H_\mathrm{kin}$ describes the electrons hopping at a rate $t_F$, $H_\mathrm{nuc}$ represents the nuclear attraction when the nuclei are at positions $\rr_n$, while $H_{e-e}$ accounts for electron-electron repulsion. In both cases, the attractive/repulsive potential has the standard Coulomb form, $V(r)=V_0/r$.  The connection of the length/energy scales of the discrete Hamiltonian $H_\mathrm{qc}$ and the continuum one is given by:
\begin{equation}
\label{eq:correspondance}
a_0/a=2t_F/V_0 \quad \text{   and   } \quad Ry=V_0^2/(4t_F)\,,
\end{equation}
where $a_0$, $a$ and $R_y$ are the Bohr radius, lattice spacing, and Rydberg energy, respectively. Thus, we work in a regime,
\begin{align*}
&\textrm{(a) } 1\ll 2t_F/V_0 \ll N/N_e^{1/3}\,, \nonumber
\end{align*}
such that the first inequality prevents discretization effects, and the second guarantees the molecule fits in the volume of the simulator.

Our simulator then requires three ingredients (see Fig.~1a): i) cold spin-polarized fermionic atoms hopping in a 3D optical potential with a tunable tunneling rate, $J_F$, to play the role of electrons~\cite{bloch08a}. We consider spinless fermions; the spin degree of freedom can be included using an extra internal level~\cite{daley2008quantum} (see Methods). ii) Additional potentials to emulate the nuclei attraction. Since this is a single-particle Hamiltonian, it can be created through optical Stark-Shifts with an adequate spatial modulation. For example, one can use holographic techniques~\cite{sturm17a,Barredo2017b} with judiciously optimized phase masks to engineer a Coulomb-like spatial potential at the fermionic positions (see Methods). iii) 
The most difficult part is to simulate $H_{e-e}$, since it involves repulsive interactions between the fermions with a $1/r$ dependence. Inspired by how virtual photons mediate Coulomb interactions in QED, we use a spin excitation of another atomic species forming a Mott insulator to mediate the Coulomb forces between fermions (see Fig.~1b).  It is composed of $N_M^3$ atoms trapped in an optical potential with the same spacing as the fermions, and with two additional internal atomic states, $|a\rangle$ and $|b\rangle$, which describe spin excitations. Spin excitations in $a$ state interact repulsively and locally with the fermionic atoms, with strength $U$, and propagate through the long-range couplings induced by a cavity mode, with rate $J_c$~\cite{ritsch13a,Schreppler2014,Brennecke2008,Brennecke2007,Domokos2002}. The $|b\rangle$ internal state is subject to a different optical potential, such that its itinerant spin excitation propagates through standard nearest-neighbor exchange~\cite{Riegger2018}, at rate $J$. Furthermore, an external field (Raman laser or a RF field) 
drives the $|a\rangle-|b\rangle$ transition with coupling strength $g$, and detuning $\Delta$. The complete simulator Hamiltonian after the elimination of the cavity mode reads $H_\mathrm{sim}=H_\mathrm{kin}+H_\mathrm{nuc}+H_M$, with 
\begin{equation}
\begin{split}
\label{eq:bosonLat}
H_M= & \; \Delta \sum_{\jj} b_{\jj}^{\dagger }b_{\jj}+ J \sum_{\langle \ii,\jj\rangle} b_{\ii}^{\dagger }b_{\jj}+  J_c/N^3_M \sum_{\ii,\jj} a_{\ii}^{\dagger }a_{\jj}\\
&+U\sum_{\jj}a_{\jj}^{\dagger}a_{\jj}f_{\jj}^{\dagger }f_{\jj} +g\sum_{\jj}\pa{a_{\jj}^{\dagger }b_{\jj} + b_{\jj}^{\dagger }a_{\jj}}  \,,
\end{split}
\end{equation}
being $a_\jj/b_\jj$ annihilation operators for a $|a\rangle/|b\rangle$-spin excitation in site $\jj$.  Intuitively, the on-site interaction $U$ localizes the $|a\rangle/|b\rangle$ excitations around the fermions, renormalizing their tunneling rates and creating an effective interaction. Mathematically, one can adiabatically eliminate the Mott insulator excitations and derive the effective dynamics for the fermions (see Supplementary Information section 2). The fermionic part of the simulator Hamiltonian $H_\mathrm{sym}\approx H_\mathrm{qc}$, with $t_F=J_F(N_e-1)/N_e$, where the electron-electron potential follows a Yukawa form~\cite{DeVega2008a} with a constant energy shift $C$:
\begin{align}
 V(r)&\approx C+\frac{V_0}{r/a}e^{-r/L}\,,
\label{eq:yukawa}
\end{align}
where $L/a=\sqrt{J/(U+\Delta+\rho_M J_c-6J)}$ is the localization length, which can be tuned with $\Delta$, and $V_0=\frac{g^2}{2\pi J N_e}$ the strength of the potential repulsion. Here $\rho_M=N_e/N_M^3$. This mapping between $H_\mathrm{sym}$ and $ H_\mathrm{qc}$ holds as long as,
\begin{align*}
 &\textrm{(b) } J_c  \ll U\,, \\
 \textrm{ (c) }J_F\ll J_c \rho_M &\sqrt{N_e}, \, \text{ and } V_0 \sqrt{N_e}  \ll J_c  \rho_M\,\,, \\
 \textrm{ (d) }  V&_0 \, N_e^2\ll J N\,(a/L)^2\,.
\end{align*}

\begin{figure}[!htb]
\centering
\includegraphics[width=\linewidth]{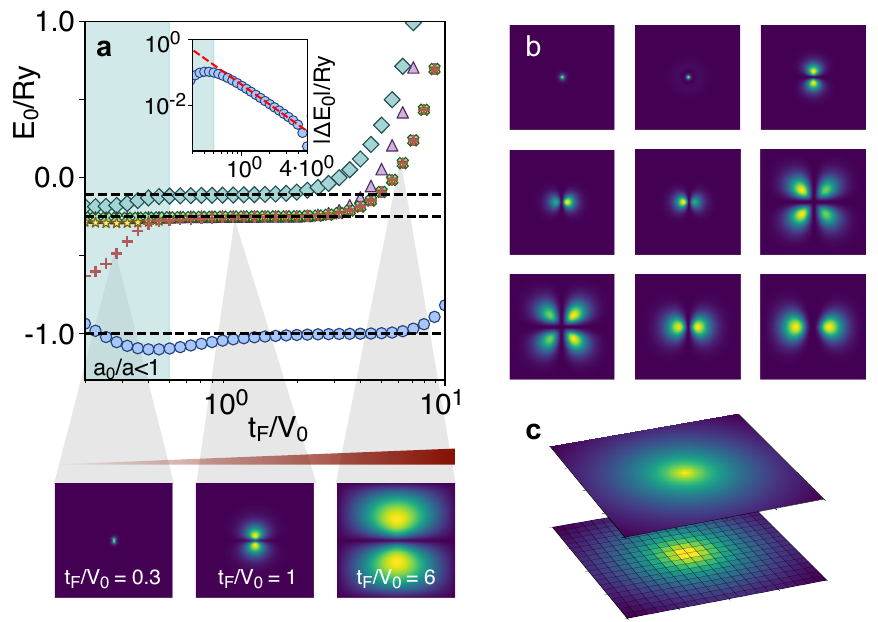}
\captionsetup{labelformat=empty}
\caption*{\textbf{Fig. 2 $|$ Atomic Hydrogen spectrum dependence on the effective Bohr radius.} \textbf{a,} Lower part of the spectrum of the atomic Hamiltonian $H_\mathrm{qc}$ for a cubic lattice of $N=100$. Dashed lines indicate the position of the 3 first atomic levels associated to the continuum Hamiltonian. In the colored region, $t_F/V_0 < 0.5$, the Bohr radius is smaller than the lattice spacing and energies are highly affected by the cut-off of the nuclear potential. As the hopping parameter $t_F/V_0$ increases, the system is effectively zoomed-in, as we show in the panels for the electron density of the second-lowest energy orbital. This includes more lattice sites in the simulation, reducing systematic deviations as $(t_F/V_0)^{-2}$ (red dashed line), as we show in the inset for the lowest energy state (see Supplementary Information section 1). At higher values of $t_F/V_0$, solutions suffer from finite-size effects. \textbf{b,} Axial cut in the central positions of the lattice is represented for the first 9 eigenstates of $H_\mathrm{qc}$ for $t_F/V_0=2$, $N=150$.
 \textbf{c,} Appropriately choosing the Bohr radius, we show how the same orbital can be obtained with $N=1000$ (up, $t_F/V_0=150$) or $N=20$ (down, $t_F/V_0=3$), where the discretization of the system is more noticeable.}
\end{figure}

\begin{figure*}[!htb]
\centering
\captionsetup{labelformat=empty}
\includegraphics[width=0.7\linewidth]{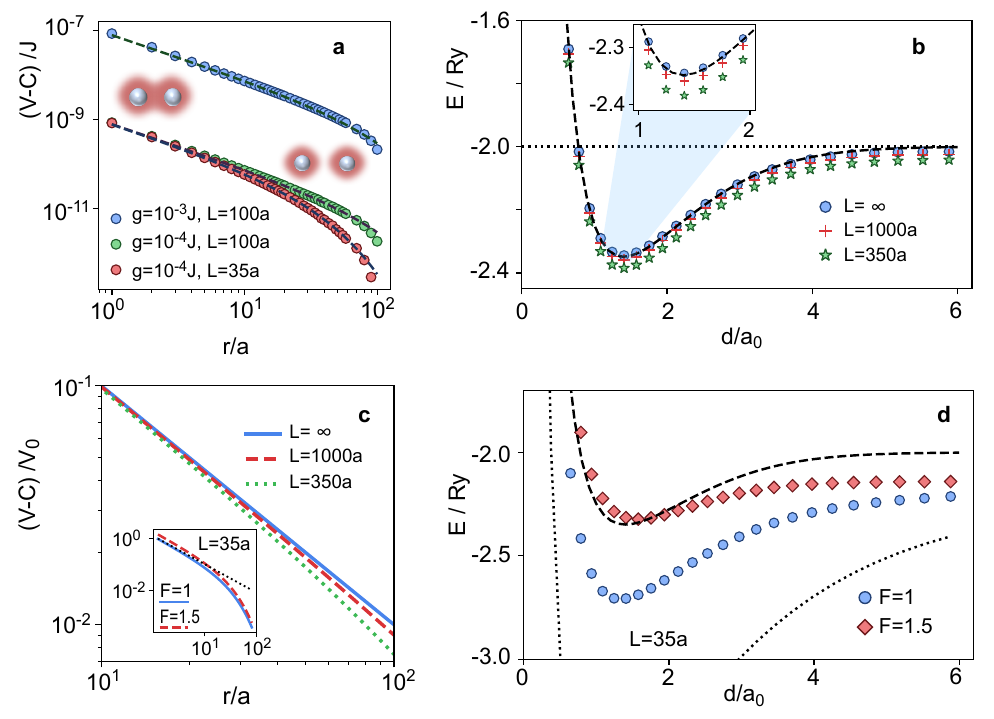}
\caption*{\textbf{Fig. 3 $|$ Molecular potential and effective interaction mediated by the Mott insulator.} \textbf{a,} Energy of the single-excitation bound state of Hamiltonian \eqref{eq:bosonLat} for two fixed fermions as a function of their separation, $r$. We choose $\Delta=2J,\, N_M=200, \, J_c=J$, such that (a-e) inequalities are satisfied. The Yukawa potential of Eq.~\eqref{eq:yukawa} corresponding to each configuration of parameters is plotted with dashed lines. \textbf{b,} We use this effective interaction to calculate the molecular potential associated to an analog simulator of $H_2$ of size $N=75$. For each internuclear separation, we choose $t_F/V_0$ giving optimal accuracy (see Supplementary Information section 3), ranging from $t_F/V_0=4.2$ to $t_F/V_0=2.3$ in the dissociation limit (dotted line). Molecular orbitals are included in the projective basis until convergence is observed. For a Coulomb potential (blue dots), the result agrees with an accurate solution in the nonrelativistic regime \cite{Kol1964,Sims2006} (dashed line). As $L$
decreases, the exponential decay in the Yukawa potential prevails, underestimating Coulomb repulsion and lowering the molecular potential. \textbf{c,} This underestimation of the repulsive potential is stressed when violating $N\ll L/a$ (see inset).  \textbf{d}, Changing the ratio, $F$  between the electronic and nuclear potential, one can explore artificial repulsive interactions that form pseudomolecules in more relaxed experimental conditions. The dotted line here represents the limit of zero-repulsion in the absence of a mediating excitation.}
\end{figure*}

Condition (b) enforces that the $a$ excitation localizes symmetrically only around the position of the fermions; (c) guarantees that neither the tunneling of the fermions nor the interaction with the $b$-excitations spoils the effective interaction; (d) ensures that the Yukawa potential does not depend on the fermionic positions. Furthermore, to obtain a truly Coulomb repulsion, the length $L$ must be larger than the fermionic lattice $N$, but smaller than the Mott insulator size, that is:
\begin{align}
 &\textrm{(e) } N\ll L/a <  N_M.\nonumber
 \end{align}
 
When all (a-e) inequalities are satisfied, the exact solution in the continuum limit is recovered in the limit $N_M>N\rightarrow \infty$. Thus, the finite size of the simulator is what ultimately limits the precision of the simulation. 

We now benchmark our simulator for moderate system sizes using numerical simulations. In Fig.~2 we solve the Hydrogen problem in a lattice to explore discretization and finite size effects by comparing the energies of the low excited states with that of the continuum. We show that an error of $0.3\%$ with respect to the exact energy can be obtained for systems of $N=100$. In Fig.~3 we analyze the accuracy for the simplest molecule, $H_2$. First, we compute exactly the energy of the spin excitation that mediates the fermionic repulsion, as a function of the interfermionic separation (Fig. ~3a). We show that it reproduces the $1/r$ behavior over a wide range of values of $g/J$ and $L$. In Fig.~3b we compute the molecular potential with $N=75$ by using a Yukawa electronic potential with different lengths $L$. We observe excellent qualitative agreement for all $L$'s considered in the figure, and a quantitative matching when $L\gg aN$. Remarkably, even if $L\lesssim aN$, valuable information can still be extracted by adjusting other experimental parameters. In Fig.~3d, we illustrate how one can increase the $V_0$ of the electron repulsion to compensate the underestimation of the potential at long distances, and obtain a pseudomolecular potential that is qualitatively similar to the one expected with Coulomb interactions.

We consider some practical issues for the implementation of these ideas (see Methods). As atomic species fulfilling the requirements we propose to use two isotopes of alkali-earth atoms, namely $^{87}Sr$ and $^{88}Sr$ for the fermions and bosons, respectively. The quantum simulator can be initialized by using adiabatic preparation, where the hopping, nuclear attraction, and interactions are sequentially turned on. We also propose to read out the physical properties by measuring the total energy of the system, by detecting the kinetic energy of the fermionic atoms under different conditions and repeating the measurements. In the Methods section we also analyze some sources of errors, and ways to circumvent them. We emphasize that some of the elements and conditions required in this proposal are beyond the state of the art. However, the rapid progress of analog quantum simulation may well lead to the realization of the present ideas in the near future, motivated by its potential impact in the determination of chemical structures, the understanding of reaction mechanisms, or the development of molecular electronics. Furthermore, with judicious changes in the implementation, e.g., different optical potential geometries, conditions (b-d) may be relaxed. Thus, we believe our proposal is a promising complementary approach to the fault-tolerant quantum computer required to solve the same problems.

In summary, we have shown how to simulate quantum chemistry problems using cold atoms in optical lattices embedded in a cavity. We expect the present proposal to stimulate both theoretical and experimental research, even before the realization of a fully-fledged analog simulator for quantum chemistry. Fig. 4 provides a roadmap of experiments with increasing complexity (as moving to the bottom right of the table) towards complete analog chemistry simulation. For instance, first experiments could be performed with spin-less fermions in one and two spatial dimensions. Another simplification might come from non-Coulomb nuclear potentials, e.g., in form of a gaussian, which does not require holographic techniques; or by using simpler schemes to obtain the fermionic potential, e.g., using a single boson instead of a Mott insulator, and without a cavity. The latter still provides an effective repulsion between two fermions.  With these simplifications, there is a clear pathway from state-of-the-art setups towards more challenging experimental setups based on progress in technology. Most important, in all these intermediate proof-of-principle setups one could already observe molecular-like potentials, dissociation, and other basic phenomena in chemistry. Besides, such experiments can also prove valuable to benchmark various numerical techniques, and trigger the development of new theoretical methods, and thus reaching a deeper understanding of the problems that appear in chemistry and which are challenging to test with classical computers.

\begin{figure}[tbp]
\centering
\captionsetup{labelformat=empty}
\includegraphics[width=\linewidth]{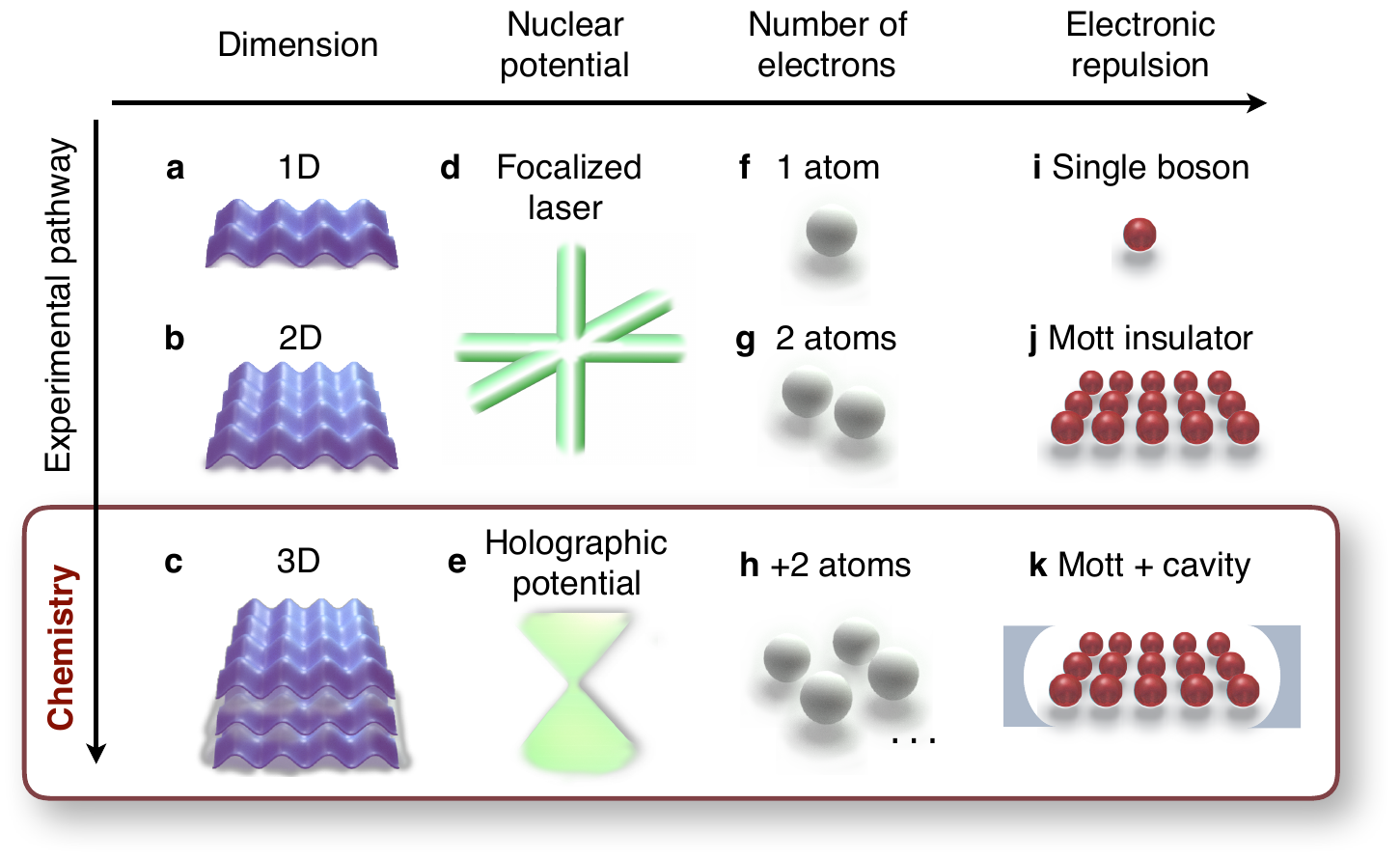}
  \caption*{\textbf{Fig. 4 $|$ Experimental pathway}. Schematic representation of the simplifications that can be considered for the different interactions of the system. The lowest line \textbf{c}+\textbf{e}+\textbf{h}+\textbf{k} corresponds to the full proposal in this work, leading to the chemical interactions we observe in nature. Chemistry in lower dimensions could be considered  by restricting the optical lattice in such directions (see \textbf{a}+\textbf{b}). The holographically created Coulomb potential could be replaced by, e.g., the Gaussian profile of a focalized laser (see \textbf{d}), giving a different scaling for electron-nuclei attraction. First implementations with single atoms (see \textbf{f}) would allow to observe simple electronic orbitals (such as the energy levels of Hydrogen) with no need to mediate repulsion. For only two atoms (see \textbf{g}) one doesn't need the symmetrizing effect of the cavity when mediating the Coulomb repulsion with the Mott insulator (see \textbf{j}). Different scalings for repulsion can also be explored in more simplified setups, such as using a single boson that hops on the lattice and has on-site interaction with the fermionic atoms (see \textbf{i}).}
\end{figure}

\section*{Acknowledgments}
\vspace*{-0.1in}
We thank S. Blatt and N. Meyer for useful discussions regarding the experimental implementation of this proposal. The authors acknowledge the ERC Advanced Grant QENOCOBA under the EU Horizon 2020 program (grant
agreement 742102). J.A.-L. acknowledges support from "la Caixa" Foundation (ID 100010434) under the fellowship LCF/BQ/ES18/11670016. Work at ICFO is supported from the Spanish Ministry of Economy and Competitiveness through the "Severo Ochoa" program, Fundaci\'o Privada Cellex, and from Generalitat de Catalunya through the CERCA program. A. G.-T. acknowledges support from the Spanish project PGC2018-094792-B-100 (MCIU/AEI/FEDER, EU) and from the CSIC Research Platform on Quantum Technologies PTI-001. T.S. is grateful to the Thousand-Youth-Talent Program of China. P.Z. was supported by the ERC Synergy Grant UQUAM, and SFB FoQus by the Austrian Science Fundation.

\newpage
\section*{Author contributions}
\vspace*{-0.1in}
J.A.-L., A.G.-T. and T.S. designed the model, developed the methods, and analyzed the data. J.A.-L. and A.G.-T. performed the calculations.
J.A.-L., A.G.-T. and J.I.C. wrote the manuscript with input from all authors. J.I.C. and P.Z. conceived the study and were in charge of overall direction and planning.

\section*{Methods}
Here we provide further details and candidates for the experimental implementation of some of the key ingredients of this proposal.

\subsection*{Holographic engineering of the nuclear potential}
Holographic techniques are a candidate approach to engineer the three-dimensional Coulomb potential seen by the fermionic atoms, $H_{\text{nuc}}$. Following this method, the potential is created by imprinting a phase pattern on an incoming laser, as used in Ref. \cite{Barredo2017b}. There, it allows to experimentally induce microtrap potentials for single atoms, and the same technique can engineer more general intensity patterns at the lattice sites where the fermions are trapped by choosing an appropriate phase mask. For this, we have particularized the algorithm used in that implementation (Gerchberg-Saxton algorithm ~\cite{Gerchberg1971}, G-S) to identify the phase mask leading to the particular geometry needed in our proposal (a 3D Coulomb potential, $V_\jj=V_0/\|\jj-\nn \|$ ).

We then use the ping-pong strategy described by the discretized G-S algorithm to retrieve the phases of the holographic field $u$ associated to our intensity distribution ($\abs{u_j}^2=V_\jj$). This approach iterates over the following steps~\cite{Chen2005}:
(i) The field is transformed to reciprocal space, obtaining $\tilde{u}=\text{FFT}(u)$. 
(ii) We restrict to those $\kk$-vectors that satisfy the physical constraints, i.e. they sit on the Ewald sphere of radius $k_0=2\pi/\lambda$, being $\lambda$ the wavelength of the monochromatic input light. The rest of contributions are neglected, defining a constrained field in Fourier space, $\tilde{u}^c$.
(iii) We transform back to real space, $u^c=\text{IFFT}(\tilde{u}^c)$. 
At this point, $u^c$ satisfies diffraction laws, but may differ from the goal potential. (iv) The objective potential is combined with the phases of the constrained solution, $\varphi_j=\text{arg}\pa{u^c_\jj}$, giving $u_\jj=\sqrt{V_\jj} e^{i\varphi_\jj}$. To improve the accuracy, we use a refined phase mask of size $\pa{n_{div}\cdot N }\times \pa{n_{div}\cdot N}$, being $N$ the number of lattice sites per side, and $n_{div}$ a refining factor (see insets in the Extended Data Fig. 1). As the trapped atoms are only affected by the value of the holographic potential on the lattice sites, these are the only coordinates where the field is updated in step (iv).

In the Extended Data Fig. 1 we show the result of applying such optimization process to the Coulomb potential. By quantifying a normalized relative error~\cite{soifer2014iteractive},
\begin{equation}
\label{eq:err}
\epsilon = \frac{1}{N^3} \sum_\jj \pa{\sqrt{V^*_\jj}-\sqrt{V_\jj}}^2/\abs{V_\jj}\,,
\end{equation}
between the desired Coulomb potential ($V_\jj$) and the intensity pattern numerically obtained ($V^*_\jj=\abs{u_\jj}^2$), we observe that the accuracy obtained for $n_{div}=3$ already provides normalized relative errors smaller than $0.3\%$ for $N=30$. More precise results can be obtained by increasing the iterations of the algorithm or the refining factor. 

\subsection*{Candidate atomic species}
This proposal is based on the interplay between two atomic species: (i) Fermionic atoms: they have two internal levels, that play the role of the electronic spin. (ii) Mediating atoms: they must have three levels available: $|0\rangle$ for the ground state in the Mott insulator; $|a\rangle$ with the spin excitation; $|b\rangle$ with the state that tunnels and induces the effective repulsion.

All atoms need to see the same lattice, although with tunable tunneling amplitudes. The fermions must additionally see the external potential and have to interact with the internal state of the mediating atoms. The scattering lenghts for the interactions corresponding to levels $|0\rangle$, $|a\rangle$ and $|b\rangle$ do not need to be the same, and we require that the scattering length between the fermions and the mediating atom in state $|b\rangle$ is neglectful, so that the scaling of the moving excitation is unchanged. Additionally, the mediating atoms in $|a\rangle$ have to be exposed to the cavity mode.  
Also, note that the atoms that mediate the fermionic interactions can correspond to both bosons and fermions, as there is just a single excitation. 

Over the last years, many atomic species have been trapped, condensed, and used in experiments with optical lattices. For illustration, let us give a particular example based on fermionic and bosonic alkaline-earth atoms. These atoms offer a rich internal structure, with long-lived excited metastable states $^3P_0$ and  $^3P_2$.

As a particular choice, we can use $^{87}$Sr as a simulator for electrons. Its nuclear spin is $I=9/2$ and, similarly to~\cite{daley2008quantum}, one can encode the spin of the simulated electron into nuclear states, $\ket{\uparrow} \equiv \ket{^1S_0, m_I=-9/2}$ and $\ket{\downarrow} \equiv \ket{^1S_0, m_I=-7/2}$. This information is therefore protected from the electronic transitions used in the rest of the proposal. 
    
    One can now use one of its bosonic isotopes, $^{88}$Sr for the mediating atoms. We assign the long-lived states $^1S_0$,  $^3P_0$, and  $^3P_2$ as levels $|0\rangle $, $|a\rangle $, and $|b\rangle $, respectively. Since there exists a magic wavelength that makes the trapping of the states $S$ and one of the $P$ equal, it is possible to choose nearby frequencies that provide basically the same lattice period for all states with the appropriate conditions. Apart from that, an additional laser driving the  $^1S_0$ level off resonance can be used to induce the external potential (with the holographic techniques explained above). Note that the isotope shift is sufficiently large in those atoms so that this can be negligible even in the case that the level  $^1S_0$ is chosen as $|0\rangle $. The cavity can be tuned close to the  $^3P_0 \to ^3S_1$ resonance, without affecting the other states. Two-photon Raman transitions can then couple the levels  $^3P_0$ and  $^3P_2$ appropriately, or one can also use a four-photon transition using level  $^1S_0$. Also, note that having nuclear spin 0 is not a problem in this case, as the mediated electronic repulsion is spin-independent. By choosing an isotope of Sr, an optical lattice with the same spacing can be engineered at the bosonic state $^1S_0$, with similar choices to the fermionic atoms. Furthermore, the lattice depth and spacing in $^3P_2$ can be independently controlled by, e.g., using magic wavelengths \cite{ye2008quantum}, and its scattering length can be tuned using, e.g., non-resonant light \cite{crubellier2017controlling}.  
   
\subsection*{Adiabatic preparation} 
A crucial step is the preparation of the appropriate states of $H_\text{sim}$, which needs to be carefully designed for the molecular system one wants to explore. In general, one is interested in the ground state properties of the electronic configuration for given positions of the nuclei. In this line, one option consists on sequentially adding the different interactions involved in the simulation.

As a particular example, let us illustrate this approach for the ground state of $H_2$. It corresponds to a singlet, electrons then have opposite spin, and are therefore distinguishable. We propose the following strategy. First, we create a product state of tightly trapped fermions that respects the  symmetries of the problem. Second, we slowly allow fermions to hop. The attraction to the nuclear potential then leads to single particle orbitals. Finally, we adiabatically introduce the electronic repulsion, completing the entire Hamiltonian.  

As an initial state, we begin with both fermions tightly trapped in a lattice site $\jj_0$ ($t_F(0)=0$). Using an external laser we create a single Mott excitation in level $a$ at this position. The initial on-site interaction  $U\neq 0$ with the fermionic atoms maintains this excitations at $\jj_0$. We can define this state as $\ket{\psi_0}=f^\dagger_{\uparrow,\jj_0}f^\dagger_{\downarrow,\jj_0} a^\dagger _{\jj_0} |0\rangle$, where $f^\dagger_{\uparrow/\downarrow,\nn}$ denotes the creations operator of one spin up/down fermionic atom in position $\nn$, and $a^\dagger_{\nn}$ the creation of a single excitation of the Mott insulator in that site. One can now adiabatically evolve this initial state according to the following steps.

\begin{enumerate}
\item Both the fermionic atoms and the Mott excitations are coherently transported into a symmetric superposition of two positions, $\nn_{1/2}$, where the nuclear potentials will be centered, separated by the desired number of sites $d/a$. This is, a new state $\ket{\psi_I}=2^{-3/2}(f^\dagger_{\uparrow,\nn_1}+f^\dagger_{\uparrow,\nn_2})(f^\dagger_{\downarrow,\nn_1}+f^\dagger_{\downarrow,\nn_2})(a^\dagger_{\nn_1}+a^\dagger_{\nn_2}) |0\rangle $. One can use different strategies for doing this, such as using moving double well potential that adiabatically transports the wavepacket in opposite directions \cite{mandel2003coherent}, or using standing waves \cite{anderlini2006controlled}. As the on-site interaction (U) is present, the Mott excitation is also transported, mediated by the long-range cavity interaction $J_C$.

At this point, the holographic Coulomb potential described in the previous sections is induced. Given that hopping processes are inhibited and fermionic atoms are now already centered in the nuclear positions, no evolution will be observed. Also, the coupling between the excited levels $a/b$ is switched-off ($g=0$). Thus, the repulsive interaction mediated by the Mott insulator is inactive. The on-site interaction $U$ modifies the fermionic hopping ($J_F \to t_F$) as detailed in the main text. Given that the optical lattice is infinitely deep ($t_F=0$) at this point, it translate into an effective null Bohr radius ($a_0(0)/a=0$). The nuclear separation is then infinite ($d/a_0=\infty$), and we have therefore prepared the dissociation limit. 

\item The next step consists on increasing the orbital size. For this, we adiabatically relax the optical lattice, slowly increasing the value of $t_F$. From the point of view of the molecular potential, this corresponds to growing the Bohr radius, therefore decreasing the effective distance $d/a_0$ (even though $d/a$ remains fixed). As only the kinetic and attractive terms of the Hamiltonian are present, there is no interaction between the two fermions, and the resulting eigenstate corresponds to two independent ground state electrons associated to $H_2^+$.

\item Once these single particle orbitals are prepared, one can adiabatically introduce the repulsive interactions. For this, we slowly couple Mott levels $|a\rangle$ and  $|b\rangle$ using a Raman transition of intensity $0\to g$. As we derived in the manuscript, this is the final ingredient required to induce an effective repulsion among the fermionic atoms, proportional to $g^2$. One can then increase $g$ until repulsion equals $V_0$.
\end{enumerate}

In the Extended Data Table I we have summarized the parameters that are modified at each stage. In the Extended Data Fig. 2 we numerically simulate this adiabatic path in 2D (and $1/r$-interactions) to prepare the ground state for a nuclear separation $d/a_0=10$, showing that such adiabatic preparation is indeed feasible within reasonable parameters. 

\subsection*{Measurement}
From the chemistry perspective, all relevant quantities can be expressed in terms of the fermionic density. This is for example the approach used in DFT methods~\cite{Parr1989}. One possibility then consists on performing a 3D spatial tomography of the $N_e$ electron and reconstruct the fermionic density. This is very complex in practice, but with gas microscope techniques (see Ref.~ \cite{omran2015microscopic}, for instance) could be feasible.

An alternative would be measuring the energy of the system. In addition to constructing molecular potentials, scanning the energy at different nuclear configurations can provide additional information, such as the value of molecular forces (Hellmann-Feynman theorem~\cite{szabo2012modern}). For this, three elements need to be simultaneously measured: the kinetic energy $\mean{K}$, the nuclear attraction $\mean{V_n}$, and the electronic repulsion $\mean{V_e}$; such that the total energy writes as $E=\mean{K}+\mean{V_n}+\mean{V_e}$. Using sudden quenches of the Hamiltonian, such contributions can be independently converted into kinetic energy. One can then perform a time of flight measurement of the fermionic atoms expelled from the lattice, using for example ionization or fluorescence techniques. As one can observe, the measured quantities will not correspond to eigenstates of the original Hamiltonian and this will give some variance in the measurement proportional to the number of fermions. One could then repeat this procedure to gain statistically significance. Once the equilibrium point of the molecular potential is identified, the procedure can also be highly simplified as only measuring $\mean{K}$ is needed to read the total energy at that point, based on the virial theorem for molecules~\cite{Gupta2015}.

\subsection*{Experimental considerations}
\label{sec:imperfections}
A reliable simulation of the quantum chemistry Hamiltonian needs that our simulator, described by Eq. 5 in the main text, satisfies a set of inequalities (a-e). We are, however, aware that there will be other experimental imperfections that may impose extra conditions and that will have to be analyzed in detail to optimize the performance of the simulation. Among the more relevant ones are:
\begin{itemize}
 \item \emph{Finite temperature}  leads to thermal fluctuations which may spoil the simulation. Thus, these fluctuations will lead to defects in the Mott insulator (see below), an may also influence the internal states of the atoms. The latter, however, can typically be very well controlled in atomic systems as we just need the atoms to be initially in a polarized state, which is reasonably easy to prepare.
 
\item \emph{Dephasing} can be originated by fluctuations in the transitions or due to magnetic fields (as internal levels are being used). This would spoil the potential of the system as a quantum simulator. However, the first effect is small in the case of microwave or Raman transitions, and the second can be controlled in the conditions already used for condensed matter simulations~\cite{choi16a,trotzky12a}.

\item \emph{Inexact fermionic filling.}  Since fermions play the role of electrons, an inexact number of fermionic atoms hopping in the lattice translates into an erroneous effective charge in the simulated molecule. These errors can be possibly post-selected by measuring the number of fermionic atoms after the simulation is performed.

\item \emph{Defects in the Mott insulator.} The absence of Mott particles in a given lattice site will locally modify the effective fermion potential. Fermions hopping to this site cannot mediate its repulsive interaction through spin-excitations, perturbing the simulated molecular orbital around this position. Importantly, the defects will not affect the potential far from the fermion such that the final performance of the simulation will scale with the density of defects rather than their number.

 \item \emph{Spatial inhomogeneities of cavity coupling.} In the simulator Hamiltonian of Eq.~(5) of the main text, we have assumed that the $|a\rangle$-excitations couple homogeneously to the cavity mode. In general, there might be some inhomogeneities that translates in a Hamiltonian:
  
 \begin{equation}
  \frac{J_c}{N_M^3}\sum_{i,j} f_{\ii,\jj}a_\ii^\dagger a_\jj\,.
 \end{equation}
 
 The fluctuations of $f_{\ii,\jj}$ around 1 will induce an extra decoherence timescale, $\Gamma_{c,\mathrm{inh}}$, that must be smaller than our simulator parameters as well. In state-of-the-art experiments, optical cavities at wavelengths of $780$ nm  and beam-waist of $\sim60$~$\mu m$ are already available~\cite{Schleier-Smith2011}, which would roughly allow for $50-100$ local minima of the standing wave sitting in a homogeneous region.
  
 \item \emph{Cavity \& atom losses.} Even though the cavity-mediated interactions are mediated by a virtual population of photons, the cavity decay introduces extra decoherence into the system due to the emission of these virtual photons. Moreover, the atomic excited states, also virtually populated, may as well decay to free space introducing losses. Thus, the cooperativity of the cavity QED system must be large to avoid both type of losses.
 
\item \emph{Three-body losses.} Since we have fermions and there can be at most one atom per lattice site, these type of losses should be small.
\end{itemize}
From these qualitative arguments, we see that most of the possible errors of the simulation are either already under control in current experiments~\cite{choi16a,trotzky12a} or scale in an intensive way.

\section*{Competing interests}
\vspace*{-0.1in}
The authors declare no competing interests.

\section*{Code and data availability}
\vspace*{-0.1in}
The computer code developed for this Letter is available upon reasonable request to the corresponding authors. All data supporting the findings of this study can be generated using the numerical methods described within the Methods and Supplementary Information. They are also available upon reasonable request.
\newpage

\begin{figure*}[tbp]
\centering
\includegraphics[width=0.8\linewidth]{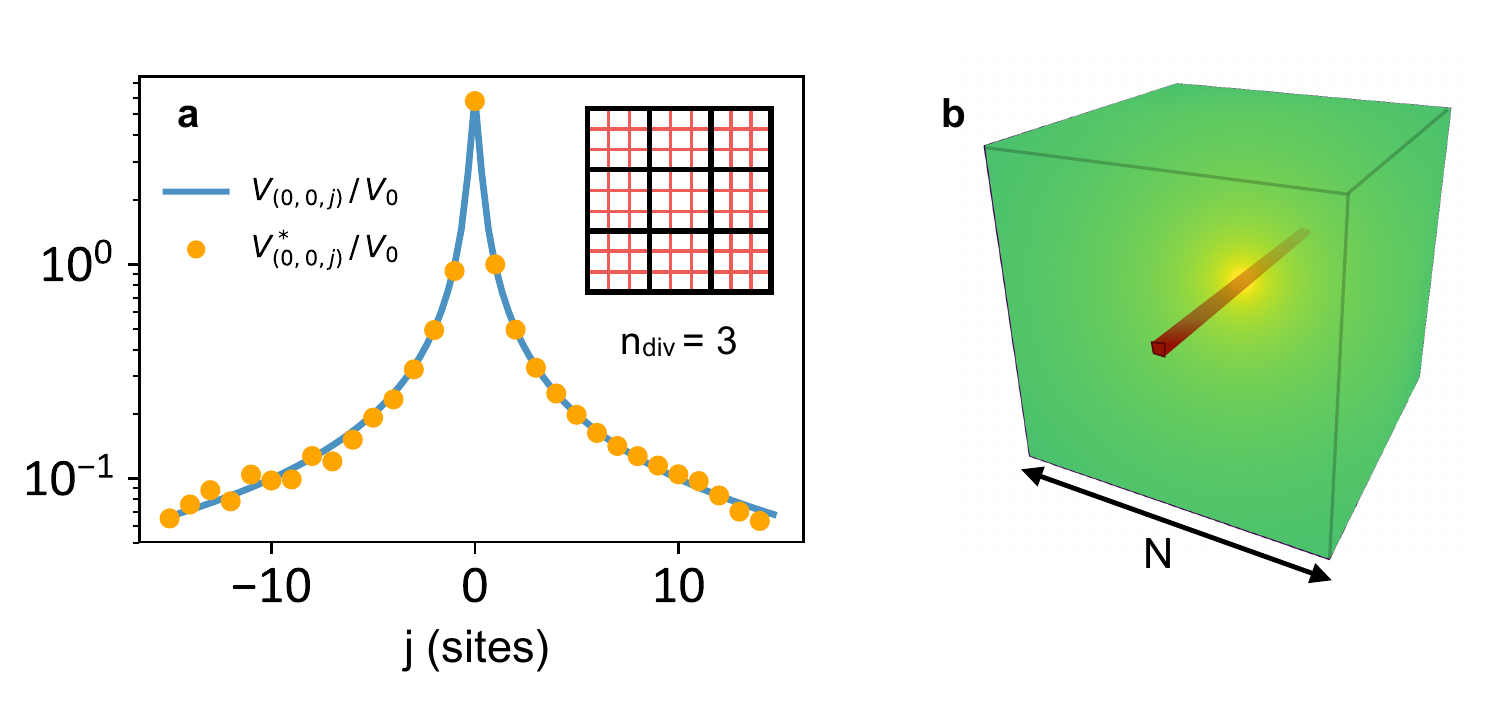}
\captionsetup{labelformat=empty}
   \caption*{\textbf{Extended Data Fig. 1 $|$ Results of the G-S algorithm}. We apply the G-S algorithm to identify the phase mask associated to a holographic 3D Coulomb potential on a lattice of $N^3$ sites. Fixing the origin at the central site, we choose the nuclear position as $\nn=[(2n_\text{ndiv})^{-1},0,0]$ (note the shift on the first coordinate, so that the lattice induces a natural cutoff). In \textbf{a} we plot an axial central cut on direction $z$ (see aligned set of red sites in \textbf{b}) of the the potential created by a phase mask composed of $(n_{div} N) \times (n_{div} N)$ cells for $n_{div}=3$ (see inset). In step (ii) of the algorithm, the Ewald sphere is discretized using a parallel projection, as used in Ref.~\cite{Chen2005}. The field is inititiated with random phases. Parameters: N=30 and 7000 iterations of the G-S algorithm.}
 \label{fig:holo_bigger}
\end{figure*}

\newpage
\begin{table}[]
\centering
\begin{tabular}{ccccc}
\toprule
 & On-site & Attractive & Fermionic  & Raman \\
Stage & interaction &  potential & hopping & coupling \\
\toprule
I(a)     & $U \to 0$    & 0       & 0                 & 0              \\
I(b)    & 0           & $V_0$        & 0                 & 0              \\
I(c)   & $0 \to U$    & $V_0$           & 0                 & 0              \\
II    & $U$          & $V_0$          & $0 \to t_F$       & 0              \\
III     & $U$          & $V_0$          & $t_F $            & $0 \to g$     \\
\toprule
\end{tabular}
  \caption*{\textbf{Extended Data Table 1 $|$ Evolution of the main parameters of the system during the adiabatic preparation} based on steps I-III presented in the Methods. To simplify the preparation of $\ket{\psi_1}$, the step I illustrated in the Extended Data Fig. 2 has been divided in three consecutive substeps:
I(a). Starting from state $\ket{\psi_0}$, the on-site interaction, $U\to 0$ is adiabatically canceled. This brings the Mott excitation into a delocalized single excitation shared by all the atoms in the insulator, $\sum_\jj a_\jj^\dagger |0\rangle /N^{3/2}$.  I(b). As proposed in the Methods, one can use a dynamic double-well potential to move the fermionic wavepackets in opposite directions. Note that, as fermionic and bosonic species are decoupled at this point ($U=0$), the Mott excitation will remain in the symmetric state reached in I(a). I(c). We adiabatically restore the on-site repulsion $U\to 0$. The bosonic state then evolves to a superposition localized at the new position of the fermionic atoms, $(a^\dagger_{\nn_1}+a^\dagger_{\nn_2}) |0\rangle /\sqrt{2}$.}
\end{table}

\begin{figure*}[tbp]
\centering
\includegraphics[width=0.8\linewidth]{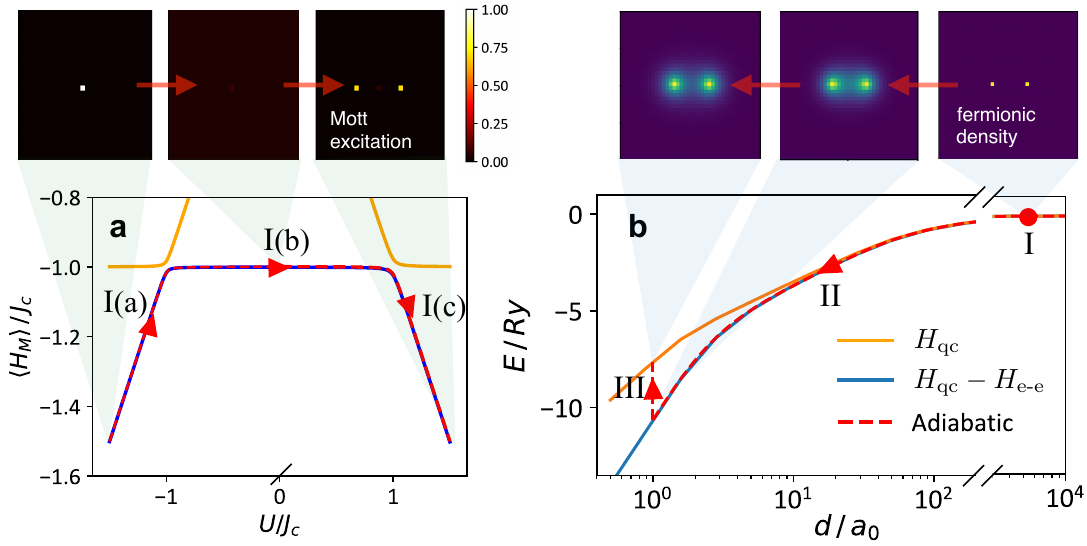}
\captionsetup{labelformat=empty}
   \caption*{\textbf{Extended Data Fig. 2 $|$ Numerical simulation of the adiabatic preparation of the ground state of $H_\text{sim}$ associated to $H_2$} (particularized for a bidimensional lattice). Red dashed lines follow the adiabatic evolution of the initial state $|\psi_0 \rangle$, and arrows point to the evolution direction. \textbf{a,} Shows the preparation of the bosonic state through steps I(a)-I(c) (see Methods and Extended Data Table. 1). Here, continuous lines indicate the exact energy of the two lowest energy sates. For the adiabatic evolution we have Trotterized time as  $\Delta t \cdot U=0.5$ and evolved with $\abs{\Delta U}/(U^2 \cdot \Delta t)=3\cdot 10^{-4}$. \textbf{b,} Continues the preparation of the fermionic state following steps (II-III). In the simulation, we have Trotterized the time evolution in intervals of $\Delta t \cdot V_0=0.05$. In (II), the kinetic term is adiabatically introduced in steps of $\Delta t_F/(V_0^2 \cdot \Delta t) =0.005$. In (III), the electronic repulsion is tuned up as $\Delta V/(V_0^2 \cdot \Delta t)=0.02$. Here yellow (blue) continuous lines follow the exact energy levels of $H_{\text{qc}}$ calculated by imaginary time evolution with(out) the effect of electronic repulsion. Panels show the Mott excitation/fermionic population in the lattice at the key moments indicated in the figure. The final point of the evolution shown in \textbf{b} corresponds to $d/a_0=1$. \emph{Parameters:} $N=60$, $U/J_c=1.5$, $d/a=10$. }
 \label{fig:adiabatic}
\end{figure*}

\newpage
\section*{Supplementary Material}
\renewcommand{\theequation}{S\arabic{equation}}
\setcounter{equation}{0} 
Here we provide details about the results and scalings stated in the main text. It is structured as follows. In Section~\ref{sec:atom}, we describe the main effect of solving the quantum chemistry problem in a discrete lattice rather than the continuum. We both explain the origin of the inequality (5),
and give the details on how to plot Fig.~2 of the main text. In Section~\ref{sec:electronrepulsion} we explain how the effective Coulomb repulsion between fermions emerge in our simulator, and discuss the origin of the inequalities (8-11) 
presented in the main text. We finally devote Section~\ref{sec:molecule} to discuss both how to optimally choose the simulator parameters to accurately obtain the $H_2$ molecular potential of Fig.~3, and the numerical method employed to obtain the data in the figure. 

\section{Discretization and finite size effects in atomic Hamiltonian \label{sec:atom}}

In this Section we focus on the effects emerging from both discretization and finite size effects. Since the effect of the fermionic repulsion is discussed in detail in sections~\ref{sec:electronrepulsion} and~\ref{sec:molecule}, here we restrict to the single particle Hamiltonian, 
\begin{equation}
H_\mathrm{kin}+H_\mathrm{nuc} =  -t_F\,\sum_{\langle \ii,\,\jj \rangle}f^\dagger_\ii f_ \jj -\sum_{n,\,\jj} Z_n V(\norm{\jj-\rr_n})\,f^\dagger_\jj f_ \jj \,.
\end{equation}

The mapping of the typical length/energy scales of the atomic problem, that are the Bohr radius ($a_0$) / Rydberg energy ($R_y$) is given in Eq.~4 of the main text. In Fig.~S1 we present a more systematic benchmarking of these identities, focusing on $1s$ orbitals of atomic Hydrogen. Since this is a quadratic Hamiltonian, we can use exact diagonalization to obtain the lowest part of the spectrum of a single electron and a single nucleus, i.e., an Hydrogen atom. This is also what we do in Fig.~2 of the main text for several ratios of $t_F/V_0$, i.e., for several expected atomic sizes. To avoid the divergence of the nuclear potential, we chose its central point as\footnote{By $(\cdot,\cdot,\cdot)$ we denote coordinates in the lattice written in units of the lattice spacing, $a$. Integer coordinates correspond to the lattice sites.} $(m,m+1/2,m)$, which induces a natural cut-off of the divergence. Here $m=\fl{N_M/2}$, where $\fl{\cdot}$ $\pa{\cl{\cdot}}$ represents the floor (ceiling) function. We also use open boundary conditions in all the figures of the manuscript. To appropriately compare the energies with the continuum limit, we shift the extracted energies by $6t_F$  and finally divide by $R_y$ to express the result in atomic units. 

As the Bohr radius increases, we observe that the numerical result approaches the analytical value in the region $1<a_0/a  < N/ N_e^{1/3}$, that is, the (5) inequality of the main text for a single electron. The first limit stems from the fact that more than one lattice site is required to properly describe an atomic orbital. The upper-bound of the inequality guarantees that the orbitals fit into the simulator volume to prevent finite-size effects, where we have assumed that the orbital size scales as $N_e^{1/3}$, based on the radial electronic density of alkali atoms. Within this range of parameters, deviations from the analytic value come from the discretization of the lattice. For instance, one can analytically compare the integral in the continuum, $\int d\rr \psi(\rr)^2 V(\rr)$ to the Riemann sum that we are effectively computing for Slater wavefunctions with radial density $2e^{-r/a_0}/(a_0)^{3/2}$. One obtains that the deviation in the energy decreases with the effective Bohr radius as $(t_F/V_0)^{-2}$, which we numerically confirm for the 
studied orbitals as illustrated in the inset of Fig.~2a. We 
observe that deviations smaller than $0.3\%$ can be reached for a lattice size $N=100$ and $t_F/V_0 < 5$.

\section{Effective fermionic repulsion \label{sec:electronrepulsion}}

Obtaining Coloumb repulsion between fermionic atoms is the crucial ingredient of our proposal. In this section, we give full details on how this effective potential emerges, and the conditions (8-11), given in the main text, that need to be satisfied to derive it. 

\subsection{Complete simulator Hamiltonian and analysis procedure}
As described in the main text, the complete simulator Hamiltonian is composed by three terms: $H_\mathrm{sim}=H_\mathrm{kin}+H_\mathrm{nuc}+H_M$, namely, the kinetic energy term corresponding to the hopping of fermionic atoms to nearest neighbour sites with rate $J_F$, the nuclear potential attraction of $H_\mathrm{nuc}$, and $H_M$, given in Eq.~5 of the main text, which contains both the Mott insulator dynamics and the interaction between the fermionic and Mott insulator atoms.

The way we analyze the problem is to use a Born-Oppenheimer approximation for the fermionic atoms, that is, assuming their timescales to be much slower than the Mott insulator atoms dynamics. Like this, we can first solve the Mott-insulator Hamiltonian for each fermionic configuration, and then, self-consistently find both the effect of the Mott-insulator in the fermion dynamics and the conditions under which this Born-Oppenheimer approximation is valid.

\begin{figure}
\centering
  \includegraphics[width=\linewidth]{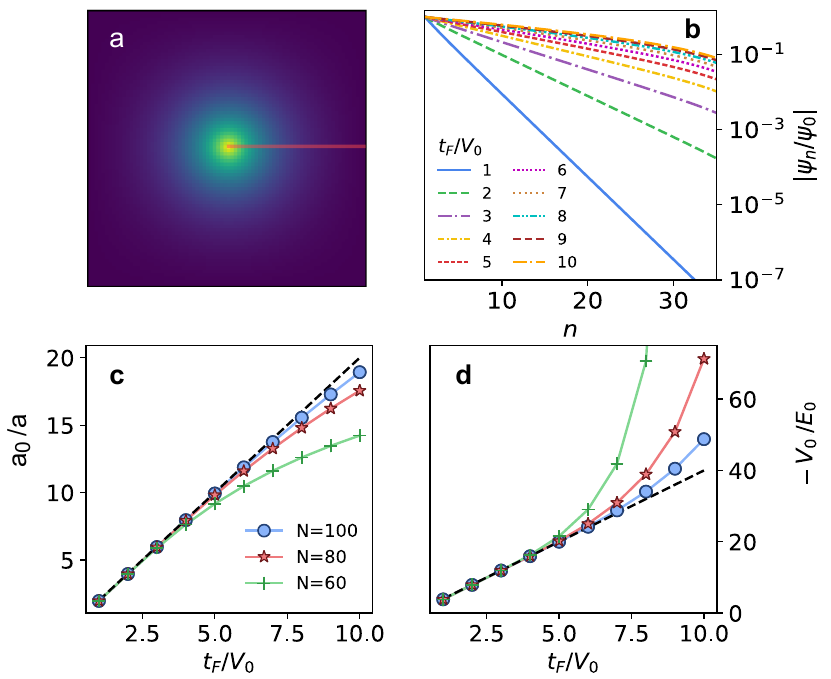}
\caption{\textbf{Validity of atomic units in finite systems.} \textbf{b,} Radial density of sates associated the ground state of $H_\mathrm{qc}$ in a cubic lattice with $N=80$ sites per side for different values of $t_F/V_0$ (in \textbf{a} we show an axial cut for $t_F/V_0=5$). After normalizing by the central density, log scale reveals the exponential radial decay $e^{-r/{a_0}}$ characteristic of $1s$ orbitals before finite-size effects appear. A linear fit of the first 30 values provide the decay parameter $1/a_0$. In  \textbf{c}, we plot this extracted Bohr radius for different sizes of the lattice and compare it to the scaling in the continuum, $a_0/a=2t_F/V_0$ (dashed line). The critical Bohr radius at which finite-size effects conspire with the scaling increase linearly with $N$. This effect is also appreciated in \textbf{d} when we compare the energy of this ground state to the dependence in the continuum, $Ry=V_0^2/(4t_F)$ (dashed line). }
\label{fig:apAtom}
\end{figure}

Thus, for a given configuration of the fermionic atoms, denoted as, $\{\jj\}=\jj_1,\dots,\jj_{N_e}$, we make the following decomposition of the Mott-insulator Hamiltonian  $H_M(\{\jj\})=H_0+H_1$, where each term reads as,
\begin{align}
\label{eq:bosonLatAp}
H_0=& \; U\sum_{\{\jj\}}a_{\jj}^{\dagger}a_{\jj}+\Delta \sum_{\jj} a_{\jj}^{\dagger }a_{\jj}+J\sum_{\langle \ii,\jj\rangle} b_{\ii}^{\dagger }b_{\jj} \,, \\
H_1=& \; J_c/N_M^3  \sum_{\ii,\,\jj} a_{\ii}^{\dagger }a_{\jj}+g\sum_{\jj}\pa{a_{\jj}^{\dagger }b_{\jj} + b_{\jj}^{\dagger }a_{\jj}}  \,.
\end{align}
Here, we will explicitly write the inverse dependence of the cavity coupling with its volume $(N_M^3)$.

Notice that $H_M(\{\jj\})$ conserves the total number of excitations, $\sum_{\jj}\left(a_\jj^\dagger a_\jj+b^\dagger_\jj b_\jj\right)$. Thus, if we initialize our simulator with a single excitation in Mott-insulator, the dynamics will be restricted to the single-excitation subspace formed by $\as/\bs=\{a^\dagger_\jj/b_\jj^\dagger \ket{0}\}_{\jj}$, where $\ket{0}$ is the state with no excitations in the $\as/\bs$ atomic states. The intuition is that the the on-site potential provided by the fermions at positions $\{\jj\}$ localizes the Mott-insulator excitations around them forming a bound state. As we will show, the energy of this bound state will depend on the particular fermionic configuration, $\{\jj\}$, in the same way than the Coulomb potential appearing in chemistry.

To show it, let us first analyze the structure of $H_0$ in the single-excitation subspace: on the one hand, it contains two degenerate subspaces with $N_e \co{N_M^*:= N_M^3-N_e}$ $|a\rangle$-states of energy $\Delta+U\,[\Delta]$ that we denote as $\as_U \,\co{\as\setminus\as_U}$, respectively. On the other hand, the subspace $\bs$ formed by the $|b\rangle$-excitations can be diagonalized if we impose periodic boundary conditions and define the operators $b_\kk^\dagger=(1/\sqrt{N_M^3})\sum_{\jj}e^{-i\kk\cdot\jj}b_\jj^\dagger$, which gives rise to the energy dispersion: $\omega_\kk=2J(\cos(k_x)+\cos(k_y)+\cos(k_z))$.

Now, we treat $H_1$ as a perturbation of $H_0$, focusing on the $\as_U$ subspace which is the one coupled directly to the fermions. Since this is a degenerate subspace, we must apply degenerate perturbation theory~\cite{sakurai1995modern}. This theory assumes that the perturbation, $H_1$ in our case, breaks the degeneracy in $H_0$, and starts by choosing an appropriate basis compatible with these new eigenstates. In our case, it is convenient to choose a basis for $\as$ privileging the symmetric combinations of excitations in the $\as_U$ and $\as\setminus\as_U$ subspaces, that can be generally written as:
\begin{align}
\ket{\varphi_{X}^{(0)}}=\frac{1}{\sqrt{\mathrm{dim}(X)}} \sum_{\jj \in X} a_{\jj}^\dagger \ket{0}\,.
\end{align}
where $X$ is the subspace $\as_U$ or $\as\setminus\as_U$, and $\mathrm{dim}(X)$ the dimension of this subspace. In particular, we are interested in the corrections to the symmetric state in $\as_U$, $\ket{\varphi_{\as_U}^{(0)}}$, which is the one that mediates the correct repulsive potential between the fermions. Thus, we need to calculate:

\text

\begin{align}
 E_{s,\as_U}&=E_{s,\as_U}^{(0)}+E_{s,\as_U}^{(1)}+E_{s,\as_U}^{(2)}+\dots\,,\nonumber\\
 \ket{\varphi_{\as_U}}&=\ket{\varphi_{\as_U}^{(0)}}+\ket{\varphi_{\as_U}^{(1)}}+\ket{\varphi_{\as_U}^{(2)}}+\dots\,.
\end{align}

As we will show afterwards, $E_{s,\as_U}$ will translate into an effective potential between the fermions, while $\varepsilon=1-|\braket{\varphi_{\as_U}^{(0)}}{\varphi_{\as_U}}|^2$ will be a measure on how much we deviate from the ideal dynamics. Thus, we will impose $\varepsilon\ll 1$ to derive the conditions (8-11) on the main text.

\subsection{Breaking the symmetric state degeneracy with cavity coupling}

As prescribed by degenerate perturbation theory, we calculate the perturbed energies/wavefunctions in two steps by separating the contribution of the cavity Hamiltonian $H_{1,c}=J_c/N_M^3\sum_{\ii,\jj}a_\ii^\dagger a_\jj$, from the rest of the perturbation $H_{1,g}=H_1-H_{1,c}$. The cavity Hamiltonian is enough to break the degeneracy of the symmetric states of $\as_U$ [$\as\setminus\as_U$], giving them an extra energy $\rho_M J_c$ [$(1-\rho_M)J_c$], respectively. Here, we define $\rho_M\equiv N_e/N_M^3$. However, apart from breaking the degeneracy it also couples both symmetric states with strength:
\begin{equation}
 \bra{\varphi_{\as\setminus \as_U}^{(0)}} H_{1,c} \ket{\varphi_{\as_U}^{(0)}}=\sqrt{\rho_M (1-\rho_M)}J_c\approx \sqrt{\rho_M}J_c\,,
\end{equation}
where in the last approximation we use $N_e\ll N_M^3$. Thus, in order to keep the mixing between symmetric states small, we must impose $U\gg J_c$, that is inequality (8) of the main text. This guarantees that:
\begin{align}
 E_{s,\as_U}^{(1,c)}\approx  \rho_M J_c\,,
\end{align}
and that the first correction to the eigenstate is:
\begin{align}
 \ket{\varphi_{\as_U}^{(1,c)}}\approx \frac{J_c\sqrt{\rho_M}}{U}\ket{\varphi_{\as\setminus\as_U}^{(1,c)}}\,.
\end{align}

Notice, that since $\rho_M\ll 1$, the inequality (8) of the main text already guarantees that $J_c\sqrt{\rho_M}/U\ll 1$ and bounds higher order correction in  $ E_{s,\as_U}^{(c)}$.

\subsection{Non-degenerate perturbation theory with $H_{1,g}$}

Once $H_{1,c}$ breaks the degeneracy, we can apply non-degenerate perturbation theory for the rest of the perturbation $H_{1,g}$. To second order, we find:
\begin{align}
\label{eq:effPot}
 E_{s,\as_U}^{(2,g)}=\frac{g^2}{N_e}\frac{1}{N_M^3}\sum_\kk \frac{\abs{e^{i\kk\jj_1}+\ldots+e^{i\kk\jj_{N_e}}}^2}{U+\Delta+ \rho_M J_c-\omega_\kk} \,.
\end{align}

In the continuum limit, that is, when $N_M \gg 1$, we can transform the sum into an integral and find:
\begin{equation}
\label{eq:yukawaIntegral}
\begin{split}
\frac{g^2}{N_e}\frac{1}{N_M^3}\sum_\kk \frac{e^{i\kk\rr}}{U+\Delta+\rho_M J_c-\omega_\kk} \approx \frac{V_0}{2 r/a}e^{-r/L}\,, \\
\frac{g^2}{N_M^3}\sum_\kk \frac{1}{U+\Delta+\rho_M J_c-\omega_\kk} \approx 0.25g^2/J-\frac{aN_eV_0}{2 L}\,.
\end{split}
\end{equation}
where $L/a=\sqrt{J/(U+\Delta+\rho_M J_c-6J)}$ is the effective length of the Yukawa-type interaction that appears in $E_{s,\as_U}^{(2)}$. This length can be controlled by the effective detuning between the symmetric state energy $U+\Delta+\rho_M J_c$, and the upper band-edge of $\omega_\kk$ at $6J$. Notice, that we define $V_0=g^2/(2\pi N_e J)$ as the effective repulsion strength.  

Before proceeding to compute the correction to the wavefunction, let us compare the result of the perturbative expansion of the energy with the exact solution of $E_{s,\as_U}$. The latter can be obtained in our problem by directly solving the Schr\"{o}dinger equation, $(H_0+H_1)\ket{\Psi_s}=E_{s}\ket{\Psi_s}$ just imposing that $\ket{\Psi_s}$ lies in the single-excitation subspace of $\as+\bs$. In the conditions of the text, the result is given in the following closed equation:
\begin{equation}
E_{s,\as_U}\approx U+\Delta+\rho_M J_c +\frac{g^2}{N_e}\frac{1}{N_M^3}\sum_\kk \frac{\abs{e^{i\kk\jj_1}+\ldots+e^{i\kk\jj_{N_e}}}^2}{E_{s,\as_U}-\omega_\kk}\,.
\end{equation}

By comparing both expression, we see that the second order correction introduced by coupling to the $b$-atomic states must be smaller that $U+\Delta+\rho_M J_c-6J$, which requires the following inequality: $N_e^2 V_0\ll J N(a/L)^2$. In this estimation, we are using a simplified formula for electron-electron repulsion~\cite{Parr1989}, $V_{e-e}\approx V_0 \pa{N_e -1}^{2/3}\sum_{\jj} \rho^{4/3}(\jj)\,,$ and approximate an homogeneous electron density, $\rho(\jj)\approx \rho$, for the optimal situation in which molecular orbitals fully occupy the lattice. This gives the (10) inequality given in the main text.

Finally, let us estimate how much the wavefunction $\ket{\varphi_{s,\as_U}^{(0)}}$ gets perturbed by $H_{1,g}$. To first order:
\begin{equation}
\label{eq:generalSt}
\ket{\varphi_{s,\as_U}^{(1,g)}} = \sum_{\kk}\beta_\kk \, b_\kk^\dagger \ket{0}\,,
\end{equation}
whose norm is given by, $\sum_{\kk}|\beta_\kk|^2$, which can be explicitly computed as,

 \begin{equation}
 \sum_{\kk}\abs{\frac{\bkev{0}{b_\kk H_{1,g}}{\varphi_s^{\as_U}}}{E_{s\as_U}^{(0)}-\omega_\kk}}^2=\frac{g^2}{N_e}\frac{1}{N_M^3}\sum_\kk \frac{\abs{e^{i\kk\jj_1}+\ldots+e^{i\kk\jj_{N_e}}}^2}{\pa{U+\Delta+\rho_M J_c-\omega_\kk}^2}\,.
 \end{equation}
 
This sum can be calculated taking energy derivatives in \eqref{eq:yukawaIntegral}, observing that the resulting condition is already guaranteed by imposing the most restrictive condition (10), that we derived by calculating the exact energy of the bound state.

Second order perturbation in $H_{1,g}$ for the wavefunction leads to two type of contributions:
\begin{equation}
\label{eq:generalSt}
\ket{\varphi_{s,\as_U}^{(2,g)}} = \sum_{s^\bot\as_U}\delta_{s^\bot\as_U} \ket{\varphi_{\as_U}^{s^\bot}}+\sum_{s^\bot\as \setminus \as_U}\eta_{s^\bot\as_U} \ket{\varphi_{\as \setminus \as_U}^{s^\bot}}\,,
\end{equation}
namely, the ones that couples the symmetric state to the antisymmetric ones in the $\as_U$ and $\as\setminus \as_U$ subspaces passing by $b$-states. The norm of the former, $\sum_{s^\bot\as_U}|\delta_{s^\bot\as_U}|^2$, can be approximately upper bounded by:
\begin{align}
\sum_{s^\bot}& \abs{\sum_{\kk}  \frac{\bkev{\varphi_{\as_U}^{s^\bot}}{H_1 b^\dagger_{\kk}}{0}\bkev{0}{b_{\kk} H_1}{\varphi_{\as_U}^s}}{\pa{E_{s,\as_U}^{(0)}-\omega_\kk}\pa{E_{s,\as_U}^{(0)}-E_{s^\bot,\as_U}^{(0)}}}}^2 \nonumber \lesssim \frac{N_e}{2}\pa{\frac{V_0 }{\rho_M J_c}}^2\,.
\end{align}

Thus, by imposing it is small we obtain one of the conditions (9) of the main text, i.e., $V_0\ll \rho_M J_c/\sqrt{N_e}$. The contribution of the antisymmetric state of $\as\setminus\as_U$, that is, $\sum_{s^\bot\as \setminus \as_U}|\eta_{s^\bot\as_U}|^2$ is already small provided that the previous inequalities are satisfied, such that it does not introduce any new condition.

\subsection{Effects on the fermionic dynamics}

Once we have calculated the energy of the bound state for each fermionic configuration, $\{\jj\}$, we now have to project the fermionic Hamiltonian:
\begin{equation}
H_f =  -J_F\,\sum_{\langle \ii,\,\jj \rangle}f^\dagger_\ii f_ \jj +H_\mathrm{nuc}+U\sum_{\{\jj\}}a_\jj^\dagger a_\jj f_\jj^\dagger f_\jj \,,
\end{equation}
into the basis that combines the $N_e$ fermions, with the symmetric spin excitation which mediates the atomic potential, that is:
\begin{equation}
\label{eq:totBase}
|| \jj_1\,,\ldots \jj_{N_e}\rangle\rangle = \pa{f_{\jj_1}^\dagger\ldots f_{\jj_{N_e}}^\dagger}  \ket{0} \otimes \ket{\varphi_{\jj_1...\jj_{N_e}}^s}\,,
\end{equation}
for  $\jj_1\,,\ldots \jj_{N_e}$ any configuration for the position of the $N_e$ fermions. To lowest order we obtain,
\begin{equation}
\tilde{H}_f \approx  -t_F \sum_{\langle \ii,\,\jj \rangle}f^\dagger_\ii f_ \jj+ H_\mathrm{nuc}+H_{e-e} \,,
\end{equation}
where one gets both the electron repulsion Hamiltonian, with a Yukawa potential $H_{e-e}$, and a kinetic energy term with a reduction of the tunneling rate, $J_F$, given by the Franck-Condon overlap:
\begin{equation}
 t_F=J_F\braket{\varphi^s_{\jj_1+1...\jj_{N_e}}}{\varphi^s_{\jj_1...\jj_{N_e}}}\,\approx J_F\frac{N_e-1}{N_e}\,.
\end{equation}

To self-consistently check the validity of the Born-Oppenheimer approach, one must certify that $H_f$ does not efficiently populate states which are not in the basis \eqref{eq:totBase} we use to project, since $H_f$ can take us out from the symmetric spin excitation, To first order in $H_f$, the perturbation to our basis states $|| \jj_1\,,\ldots \jj_{N_e}\rangle\rangle$ has the form
\begin{equation}
|| \jj_1\,,\ldots \jj_{N_e}\rangle\rangle^{(1)} = \sum_{s^\bot } \phi_{{\jj_1...\jj_{N_e}}}^{s^\bot} | \jj_1\,,\ldots \jj_{N_e} \rangle \otimes \ket{\varphi^{s^\bot}_{\jj_1...\jj_{N_e}}}\,,
\end{equation}

We can upper bound the perturbation by summing up over the different configurations to arrive to:
\begin{equation}
\sum_{\{\jj\}}\sum_{s^\bot } |\phi_{{\jj_1...\jj_{N_e}}}^{s^\bot}|^2 \lesssim N_{nn} \pa{\frac{J_F}{\rho_M J_c}}^2\,,
\end{equation}
where $N_{nn}$ denote the number of nearest neighbors in the lattice (6 in the case of a simple cubic geometry). Imposing that this correction is small, we obtain the other inequality (9) of the main text, that is, $J_F\ll \rho_M J_c \sqrt{N_e}$. 

Finally, there are two additional inequalities related to obtaining a truly Coulomb repulsion rather than Yukawa. On the one hand, the length of the bound state has to ideally be larger than the fermionic lattice, $L\gg aN$, while being smaller than the Mott insulator lattice, $L<aN_M$, such that the bound state does not feel the border. Combining both inequalities, we arrive to the condition (11) of the main text.

\section{Obtaining molecular potentials\label{sec:molecule}}

In this section, we first give more details on how we choose the simulator parameters to plot the figures of the manuscript, and explain the numerical methods employed for the two-electron wavefunction calculation.

\subsection{Choice of simulator parameters}

In Figure~3a we calculate the energy of the bound state of the spin excitation as a function on the interfermionic separation. We use exact diagonalization of $H_M(\{\jj\})$ for two fermions at positions $(m- \cl{r/(2a)},m,m)$ and $(m+\fl{r/(2a)},m,m)$. To calculate the constant energy shift $C$, we also obtain the energy of the spin excitation bound state when a single fermion is placed at position $(m,m,m)$, and subtract it from the energy of the two fermions. In this Figure, we choose the parameters such that the inequalities (5,8-11) are satisfied and obtain Coulomb repulsion between fermions for $N_M=200$.

To plot the molecular potential of Fig.~3b, that is, to calculate the electronic energy, $E(d)$ as a function of the internuclear distance $d$, we center the nuclear potential in positions $\rr_1=(m- \cl{d/(2a)},m+1/2,m)$ and $\rr_2=(m+\fl{d/(2a)},m+1/2,m)$ and obtain the ground state energy using the numerical methods explained in the next Section. Since two electrons are involved, the extracted energy is now shifted by $12t_F$, and finally written in atomic units. Notice also that, since we use spinless fermions, we have to restrict to the symmetric subspace of the electronic problem so that we can compare the results to those of the $H_2$ molecule, which is formed for two spins of opposite sign.

\begin{figure}[tbp]
\label{fig:apMol}
\hspace*{-0.35in}
\centering
  \includegraphics[width=\linewidth]{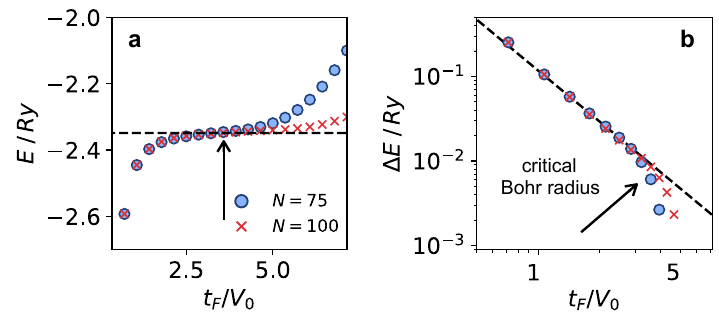}
\caption{ \textbf{Calculation of the critical Bohr radius for $H_2$.} \textbf{a,} For each internuclear separation $d/a_0$, we benchmark separations on the lattice $d/a$ ranging from 1 to 30 sites, tuning the Bohr radius, $a_0/a$, accordingly. In the figure, $d=1.4 a_0$, close to the equilibrium distance. Similarly to what happened in the atomic case (see Figure 2\textbf{a}), the calculated potential polynomially approaches the exact solution (dashed line) as the effective Bohr radius increases,  up to the point in which finite size effects appear. To calculate this optimal $t_F/V_0$, we repeat the calculation for a bigger system, and detect the point where both curves depart. \textbf{b,} To identify this point, we fit the energy of the largest lattice to a universal scaling $m\pa{t_F/V_0}^{-2}+n$, (dashed line). For $N=75$ and a nuclear separation of $1.4a_0$, the critical point (indicated by the arrow) corresponds to $t_F/V_0\approx 3.5$, providing an expected precision of $ 10^{-2}Ry$ in the energy of the minimum 
potential at this given distance. }
\end{figure}

As it happens in the atomic case, accuracy increases as the Bohr radius grows up to a critical value at which finite-size effects are relevant. However, the optimal choice of the Bohr radius now depends on the number of lattice sites that separate the nuclei. To identify this critical Bohr radius at which the finiteness of the lattice $N=75$ compromises the accuracy, we use the following procedure. First, for a given internuclear separation $d/a_0$ we solve the electronic structure for nuclear potentials separated a number of lattice sites $d/a$ ranging between $1$ and $30$, increasing the Bohr radius accordingly, $t_F/V_0=(d/a)/(2d/a_0)$. The same calculation is repeated for a bigger system size, $N=100$. Both lattices provide compatible results as long as finite-size effects are not important. The point where both curves deviate corresponds to an approximate optimal $t_F/V_0$, that provides maximum accuracy for the lattice size considered, e.g., $N=75$ in our case. In practice, we choose the point at which 
finite-size energy deviations are one order of magnitude smaller than the discretization error (see Fig. S2). Fitting these values for each internuclear separation $d/a$, we 
choose $t_F/V_0=4.2-d/a\cdot 0.065$, i.e., as the nuclei are more separated, the border of the system linearly approaches, needing to reduce the Bohr radius accordingly.

\subsection{Numerical methods to deal with the two-electron problem}

Another important difference with respect to the single electron problem is that exact diagonalization strategy is out of reach for our computational resources. To prevent this situation, we artificially reduce the degrees of freedom by projecting the Hamiltonian on a single fermion basis, $\{\phi_i\}_{i=1}^{n}$. This projected Hamiltonian reads as,
\begin{equation}
\ha_P=\sum_{i,j,r,s=1}^{n} h_{ijrs}\ket{\phi_i\;\phi_j} \bra{\phi_r\; \phi_s}\,,
\end{equation}
where $\ket{\phi_i\;\phi_j}$ denotes the product state $\ket{\phi_i}\otimes \ket{\phi_j}$ and
$  h_{ijrs}=\bra{\phi_i\; \phi_j} \ha_{\text{eff}}\ket{\phi_r \; \phi_s}\,. $

The success of this strategy now depends on how accurately the orbitals in this set can describe the interactions in the Hamiltonian. We then choose a basis composed by single-fermion states of $H_2^+$  calculated with exact diagonalization, together with more orbitals obtained using a Hartree-Fock approximation~\cite{Domokos2002}. This is, starting with the ground state obtained for a single fermion, we iterate the equation,
\begin{equation}
\label{eq:hartreeFock}
\pa{\mathcal{H}_0 \ket{\phi}}^\ii + \sum_{\jj}\pa{\phi^{\jj}}^2 V_{12}(\norm{\jj-\ii})\,\phi^\ii= \lambda \, \phi^\ii \,,
\end{equation}
until convergence is reached. This way, orbitals interact with a mean charge induced by the rest of fermionic atoms in the lattice, while we neglect the exchange interaction.

Once we have built an approximated basis, we need to project the Hamiltonian $H_\mathrm{qc}$ into the basis. The terms associated to the kinetic energy and nuclear interactions are easily projected, as they only depend on single fermionic orbitals, $\bra{\phi_i\; \phi_j} \ha_0\ket{\phi_r \; \phi_s}=\delta_{js}\bra{\phi_i} \ha_0\ket{\phi_r }+\delta_{ir}\bra{\phi_j} \ha_0\ket{\phi_s}$. The lattice imposes a natural cutoff \eqref{eq:yukawaIntegral}, corresponding to $V(0) \approx \pi V_0$, and the main difficulty comes from calculating terms associated to e-e interactions,
$\ha_{\text{ee}}$. At a first glimpse, they involve a sum of $N^6$ coordinates,
\begin{equation}
\sum_{\bb{r}_1, \bb{r}_2} V(\bb{r}_1-\bb{r}_2)\phi_i(\bb{r}_1)\phi_r(\bb{r}_1)\phi_j(\bb{r}_2)\phi_s(\bb{r}_2)\,,
\end{equation}
where, $V(\rr)=V_0/\norm{\rr}$. In the reciprocal space, however, this sum simplifies as
\begin{equation}
\bra{\phi_i\; \phi_j} \ha_{\text{ee}}\ket{\phi_r \; \phi_s}=\sum_{\bb{k}} \widetilde V(\bb{k})\cdot \widetilde{\pa{\phi_i \cdot \phi_r }}(\kk) \cdot \widetilde{\pa{\phi_j \cdot \phi_s }}(-\kk)\,,
\end{equation}
and only $N^3$ terms are involved, speeding-up the calculation. Here $\widetilde{f}$ denotes the Fourier transform of function $f$. In principle, this induces periodic boundary conditions in the system, which are undesirable as fermions would interact along the minimum distance measured on the periodic lattice, overestimating e-e interactions. To solve this issue, we double the size of the system before calculating the Fourier transform, and impose null probability of occupying these artificial positions. Fourier transforms are obtained using a Discrete Fast Fourier Algorithm. 

This last calculation is the bottleneck from the computational time perspective and, at the expense of memory, we initially store the FFT for each of the $n(n+1)/2$ product of pairs of molecular orbitals, so that the transformation is not unnecessarily repeated. It is also useful to note that not every term $h_{ijrs}$ needs to be calculated, due to the symmetries of the Hamiltonian. For example, $h_{1123}=h_{1132}=h_{2311}=h_{3211}$. In practice, this reduces the calculated terms from $n^4$ to $n^2(n^2+3)/4$ independent terms, where $n$ is the number of molecular orbitals in the projected basis. For Figure 3b, we observed that convergence in energies to the order of the discretization effects was reached for 15 Hartree-Fock orbitals and 15 low-energy $H_2^+$ states. This corresponds to $n=30$, $203175$ independent terms, and approximately 8h of total computational time in a 2.20GHz CPU.

The mean charge interaction in the Hartree-Fock calculation can also been rewritten as,
\begin{equation}
 \sum_{\jj}\pa{\phi^{\jj}}^2 V_{12}(\norm{\ii-\jj})\cdot \phi^{\ii}=\braket{\phi}{\mathcal{F}^{-1}\pa{\widetilde V(\kk)\cdot \widetilde{ \phi^2}(-\kk)}}\,,
\end{equation}
where $\mathcal{F}^{-1}$ denotes the inverse Fourier transform. We should emphasize that this projection on a single-particle basis is just a numerical strategy that enables us to numerically benchmark the model, but does not have any impact on the experimental implementation of the proposed analog simulator.

\end{document}